\documentclass[a4paper,11pt]{article}
\usepackage{pos}
\usepackage{subfig} 

\newcommand{\harpoon}{\overset{\rightharpoonup}}

\title{Associated production in pp and heavy ion collisions}

\author*[a]{Hua-Sheng Shao}

\affiliation[a]{Laboratoire de Physique Th\'eorique et Hautes Energies (LPTHE), UMR 7589, Sorbonne Universit\'e et CNRS,
4 place Jussieu, 75252 Paris Cedex 05, France}

\emailAdd{huasheng.shao@lpthe.jussieu.fr}

\abstract{Associated particle production processes in pp and heavy ion collisions at the LHC are in particular interesting in the sense that they provide unique tools to study double parton scattering (DPS) mechanism. 
In this talk, I will first review the recent theoretical, phenomenological and experimental developments of DPS in pp collisions. Then, I will focus on the DPS studies in heavy ion collisions, and stress their roles in understanding the cold nuclear matter effects, such as the (poorly known) impact-parameter dependent nuclear parton densities.}

\FullConference{%
  The Eighth Annual Conference on Large Hadron Collider Physics-LHCP2020 \\
  25-30 May, 2020\\
  online}

\begin{document}
\maketitle

\vspace{-0.5cm}
\section{Introduction}
\vspace{-0.3cm}

Associated particle production at high-energy hadron colliders, like the LHC, are unique in studying an unconventional mechanism of particle production, the so-called multi-parton interaction (MPI).  MPI is ubiquitous in proton-proton (pp) or nuclear collisions because of the compositeness nature of the incoming particles. Such MPI mechanism is indispensable in scrutinising many event activities and hadron multiplicities of observables measured at high-energy experiments. Due to the unitarity in scattering cross sections, its effect does not show up in inclusive-enough observables, such as in the single-inclusive particle production processes. On the other hand, when we are interested in the multi-particle final state that could undergo more-than-one scattering subprocesses, MPI effect in general should be taken into account. If more-than-one scattering subprocesses lie in the perturbative regime, the perturbation theory effectively works based on the factorisation approach. According to the power counting, the most viable MPI in perturbative regime is the so-called double parton scattering (DPS), in which two simultaneous hard scattering subprocesses take place in a single reaction. DPS contributes to signals and backgrounds in many analyses at the LHC.

\vspace{-0.5cm}
\section{The status of DPS in pp collisions}
\vspace{-0.3cm}

Like the conventional single parton scattering (SPS), the factorisation theorem for the double Drell-Yan process $pp\to \ell_1^+\ell_1^- \ell_2^+\ell_2^-$ has been proven recently~\cite{Diehl:2015bca,Diehl:2018wfy}. In such an factorisation approach, the lowest order (differential) cross section can be written as
\vspace{-0.3cm}
\begin{eqnarray}
&&d\sigma_{pp\to \ell_1^+\ell_1^- \ell_2^+\ell_2^-}^{{\rm DPS}}=\frac{m}{2}\sum_{i,j,k,l}{\int{dx_1dx_2dx_1^\prime dx_2^\prime d^2\harpoon{b}_1d^2\harpoon{b}_2d^2\harpoon{b}}}\nonumber\\
&&\times\Gamma_{ij}(x_1,x_2,\harpoon{b}_1,\harpoon{b}_2)d\hat{\sigma}_{ik\to \ell_1^+\ell_1^-}(x_1,x_1^\prime)d\hat{\sigma}_{jl\to \ell_2^+\ell_2^-}(x_2,x_2^\prime)\Gamma_{kl}(x_1^\prime,x_2^\prime, \harpoon{b}_1-\harpoon{b},\harpoon{b}_2-\harpoon{b}),
\end{eqnarray}
where $m$ is a combinatorial factor to account for the symmetry of the final state and $\harpoon{b},\harpoon{b}_1,\harpoon{b}_2$ are the transverse displacements of the two protons and of the partons with respect to the centres of their protons. Although the short-distance pieces $d\hat{\sigma}$ are perturbatively calculable, the unknown complex multi-dimensional objects $\Gamma$ obscure the direct applications of the formula in phenomenological studies. Several assumptions by relating $\Gamma$ to the well-known one-body parton distribution function (PDF) and the parton transverse profile simplify the above equation into a widespread ``pocket formula":
\vspace{-0.5cm}
\begin{eqnarray}
d\sigma_{pp\to \ell_1^+\ell_1^- \ell_2^+\ell_2^-}^{{\rm DPS}}&=&\frac{m}{2}\frac{d\sigma_{pp\to \ell_1^+\ell_1^-}d\sigma_{pp\to \ell_2^+\ell_2^-}}{\sigma_{{\rm eff},pp}}.
\end{eqnarray}
Such a formula has an obvious virtue that it is very predictive because the DPS cross section only relies on a single unknown number $\sigma_{{\rm eff},pp}$, which can be determined from experiment once for all. Thus, it drives most of the phenomenological studies and all the experimental measurements. It has been investigated that a few limitations of the pocket formula can be lifted by relaxing the assumptions. For instance, the scale evolution of the two-body PDF can be numerically performed via the double DGLAP (dDGLAP) equation~\cite{Gaunt:2009re,Elias:2017flu}. It has been shown in Ref.~\cite{Gaunt:2009re} that the introduction of dDGLAP evolution can lead to $10-20\%$ deviation below $x_1=x_2=10^{-1}$ with respect to the product of two one-body PDFs, while such deviations can be strongly amplified when close to the threshold $x_1+x_2=1$. Some other proposals beyond the pocket formula are present in the literature. One example is that two partons from one hadron beam could stem from a perturbative splitting of a single parton from that beam~\cite{Gaunt:2011xd,Blok:2011bu,Diehl:2011yj,Manohar:2012pe,Ryskin:2012qx}. Such a contribution, usually dubbed as $1v2$ or $2v1$, has some double counting with the dDGLAP evolution of the two-body PDF, which should be subtracted properly in a rigorous perturbative QCD computation. A second example is the introduction of parton-parton correlations~\cite{Ceccopieri:2017oqe,Cotogno:2020iio} will yield $\sigma_{{\rm eff},pp}$ to be a kinematic-dependent function instead of a number. However, such an effect is quite mild. It was anticipated to see $1\sigma$ deviation from the constant $\sigma_{{\rm eff},pp}$ at the high-luminosity LHC phase with the integrated luminosity 1 ab$^{-1}$ through the same-sign WW final state~\cite{Ceccopieri:2017oqe}.

\begin{figure}
\vspace{-1.1cm}
  \centering
  \subfloat[$\sigma_{{\rm eff},pp}$]{
    \includegraphics[width=0.48\textwidth,draft=false]{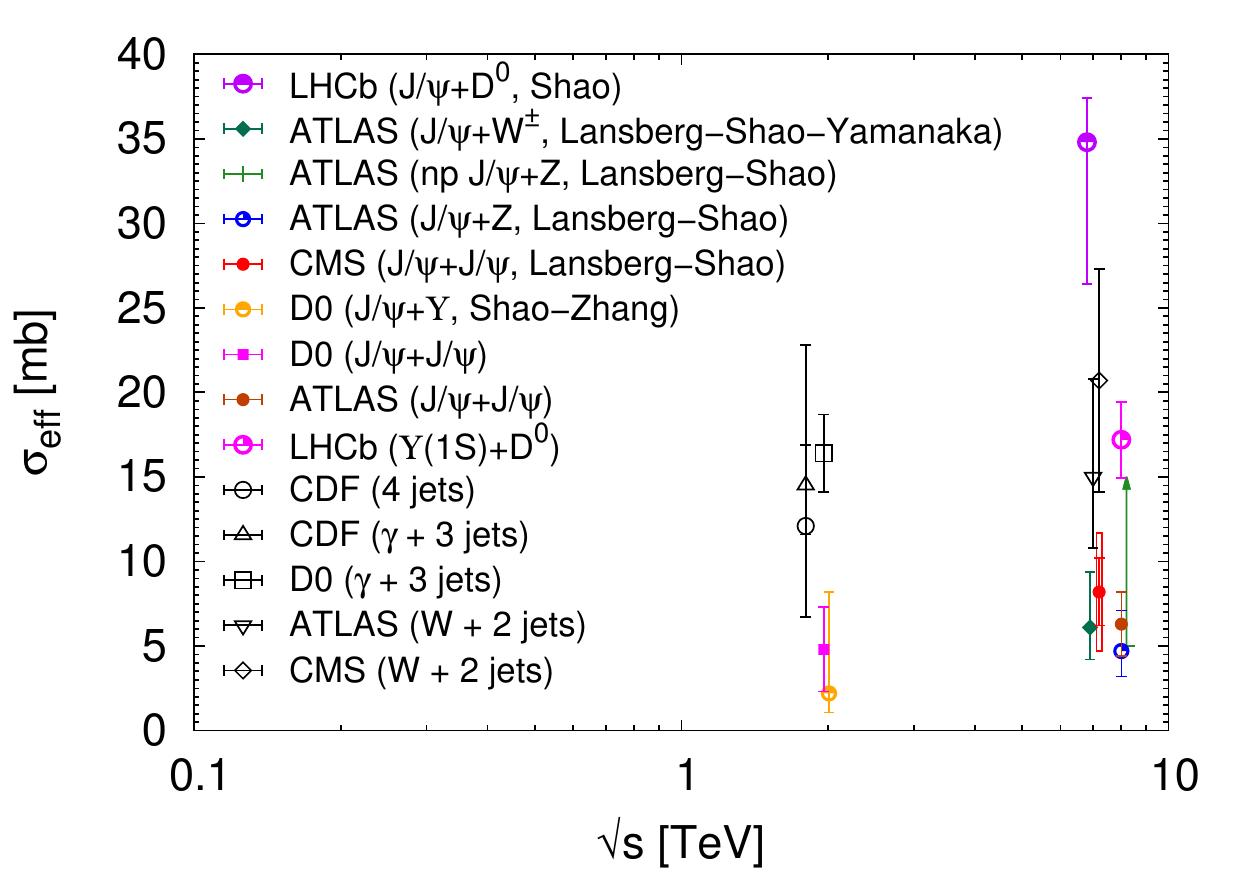}\label{fig:sigmaeff}}
  \subfloat[$R_{p{\rm Pb}}$ versus $P_T(J/\psi+D^0)$]{\includegraphics[width=0.40\textwidth,draft=false]{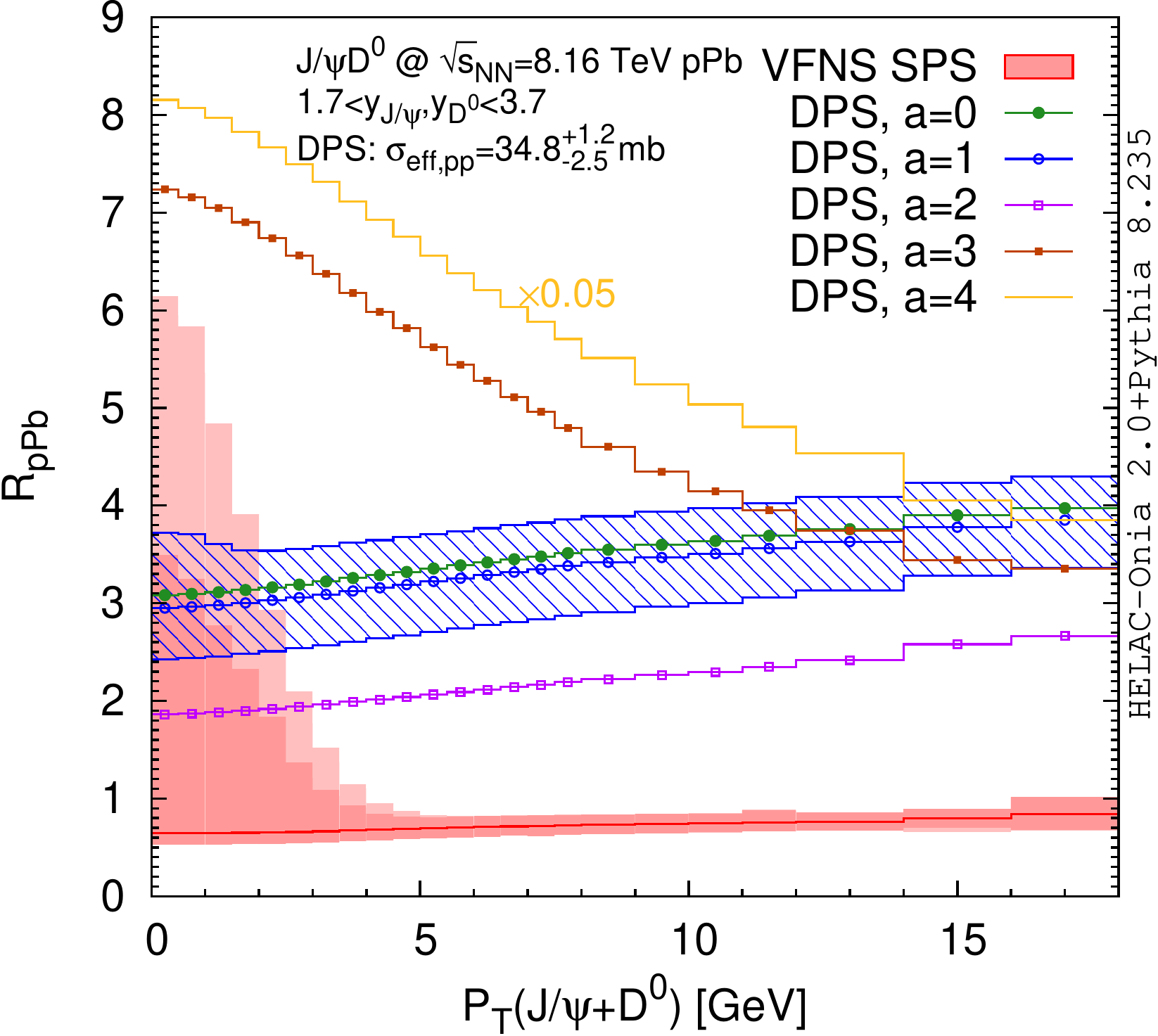}\label{fig:RpPb}}
  \vspace{-0.3cm}
  \caption{(a) $\sigma_{{\rm eff},pp}$ versus the centre-of-mass energy $\sqrt{s}$ of proton-(anti-)proton collisions; (b) Nuclear modification factor $R_{p{\rm Pb}}$ versus $P_T(J/\psi+D^0)$ for $p{\rm Pb}\to J/\psi+D^0$. The right plot is taken from Ref.~\cite{Shao:2020kgj}.
  \label{fig:plot}}
  \vspace{-0.6cm}
\end{figure}

Nonetheless, the pocket formula is a good starting point, and one can take any deviation with respect to experimental measurements as an indication of calling for a more rigorous treatment. There are existing many measurements to probe DPS or to extract $\sigma_{{\rm eff},pp}$ value from various final states at the LHC and Tevatron. The final particles generically span from jets, photons, W/Z bosons, heavy flavour hadrons and quarkonia as well as leptons. In particular, quarkonium associated production processes constitute a significant fraction (almost a half so far) of DPS studies. We have summarised $\sigma_{{\rm eff},pp}$ from various extractions (see, e.g., Refs.~\cite{Aad:2013bjm,Chatrchyan:2013xxa,Abazov:2014qba,Lansberg:2014swa,Aaboud:2016fzt,Aaij:2012dz,Aaij:2015wpa,Shao:2016wor,Lansberg:2016rcx,Lansberg:2016muq,Lansberg:2017chq,Shao:2020kgj}) in Fig.~\ref{fig:sigmaeff}. Albeit being inconclusive, it is indicative that $\sigma_{{\rm eff},pp}$ disfavours a constant with caveats that not all SPS contributions are probably well under control.

\vspace{-0.5cm}
\section{DPS studies in heavy-ion collisions}
\vspace{-0.3cm}

DPS cross sections will be enhanced more quickly than SPS cross sections in heavy-ion collisions thanks to much larger transverse parton density of heavy nuclei compared to nucleons~\cite{Strikman:2001gz,dEnterria:2012jam}. It has been recently pointed out in Ref.~\cite{Shao:2020acd} that DPS in nuclear collisions is in fact also quite useful in determining the impact-parameter-dependent nuclear PDFs (nPDFs), which are essentially unknown. Taking $p{\rm Pb}\to J/\psi+D^0$ as an example, the DPS nuclear modification factor can be written in terms of $R_{p{\rm Pb}}$'s of single particle production
\vspace{-0.2cm}
\begin{eqnarray}
&&R_{p{\rm Pb}\to J/\psi+D^0}^{{\rm DPS}}=R^{J/\psi}_{p{\rm Pb}}R^{D^0}_{p{\rm Pb}}\left[\frac{3^{1-2a}(a+3)^{2a}}{2a+3}+r_{{\rm eff},A}\frac{9^{1-a}(a+3)^{2a}}{4(a+2)}\right]\nonumber\\
&&+\left(R^{J/\psi}_{p{\rm Pb}}+R^{D^0}_{p{\rm Pb}}\right)\left[1-\frac{3^{1-2a}(a+3)^{2a}}{2a+3}+r_{{\rm eff},A}\left(\frac{3^{2-a}(a+3)^a}{2(a+4)}-\frac{9^{1-a}(a+3)^{2a}}{4(a+2)}\right)\right]\nonumber\\
&&+\left[-1+\frac{3^{1-2a}(a+3)^{2a}}{2a+3}+r_{{\rm eff},A}\left(\frac{9}{8}+\frac{9^{1-a}(a+3)^{2a}}{4(a+2)}-\frac{3^{2-a}(a+3)^a}{(a+4)}\right)\right]
\end{eqnarray}
with $r_{{\rm eff},A}\equiv \frac{\sigma_{{\rm eff},pp}}{\pi R_A^2}(A-1)\overset{A={\rm Pb}}{\simeq}5.23\left(\frac{\sigma_{{\rm eff},pp}}{34.8~{\rm mb}}\right)$, where we have used the test function $G\left(\frac{T_A(\harpoon{b})}{T_A(\harpoon{0})}\right)\propto\left(\frac{T_A(\harpoon{b})}{T_A(\harpoon{0})}\right)^a$ to characterise the spatial-dependent of nuclear modification encoded in nPDFs via $R_k^A\left(x,\harpoon{b}\right)-1=\left(R_k^A\left(x\right)-1\right)G\left(\frac{T_A(\harpoon{b})}{T_A(\harpoon{0})}\right), k=q,\bar{q},g$. $R^{J/\psi}_{p{\rm Pb}},R^{D^0}_{p{\rm Pb}}$ and $\sigma_{{\rm eff},pp}$ can be either calculable or measurable by independent experiments.

We have compared DPS and SPS $R_{p{\rm Pb}}$ of $p{\rm Pb}\to J/\psi+D^0$ in Fig.~\ref{fig:RpPb}. It turns out that $R_{p{\rm Pb}}$'s are very different between SPS and DPS. Furthermore, DPS results with different spatial dependence $a$ are also quite distinguishable. We are looking forward to the emergence of more and more DPS measurements at the LHC in the near future, while the first measurement appeared recently~\cite{Aaij:2020smi}. They will help us to understand more about the cold nuclear matter effects from different angles.

\vspace{-0.5cm}
\section{Conclusions}
\vspace{-0.3cm}

In this talk, I have discussed about taking associated particle production in pp and heavy-ion collisions as tools to deepen our understanding of DPS and cold nuclear matter effects (e.g., the impact-parameter-dependent nPDFs). I firstly review the current status of DPS studies in pp collisions from both theoretical developments and phenomenological/experimental explorations. Very impressive progress has been achieved in the LHC era. Then, I proposed to use DPS in heavy-ion collisions to reveal the impact-parameter-dependent nPDFs. Other cold nuclear matter effects in the DPS formula are also worthwhile pursuing in the future. The continuing LHC measurements on the topic will definitely allow us to learn more from various physics perspectives.

{\bf \emph{Acknowledgements.}}
This work is supported by the ILP Labex (ANR-11-IDEX-0004-02, ANR-10-LABX-63).


\bibliographystyle{Science}
\bibliography{reference}



\end{document}